# Scheduling Data Intensive Workloads through Virtualization on MapReduce based Clouds

[1]B.Thirumala Rao, [2]L.S.S.Reddy
Department of Computer Science and Engineering,
Lakireddy Bali Reddy College of Engineering, Mylavaram
[1]thirumail@yahoo.com, [2]director@lbrce.ac.in

## ABSTRACT

*MapReduce has become a popular programming model for running data intensive applications on the cloud. Completion time goals or deadlines of MapReduce jobs set by users are becoming crucial in existing cloud-based data processing environments like Hadoop. There is a conflict between the scheduling MR jobs to meet deadlines and "data locality" (assigning tasks to nodes that contain their input data). To meet the deadline a task may be scheduled on a node without local input data for that task causing expensive data transfer from a remote node. In this paper, a novel scheduler is proposed to address the above problem which is primarily based on the dynamic resource reconfiguration approach. It has two components: 1) Resource Predictor: which dynamically determines the required number of Map/Reduce slots for every job to meet completion time guarantee; 2) Resource Reconfigurator: that adjusts the CPU resources while not violating completion time goals of the users by dynamically increasing or decreasing individual VMs to maximize data locality and also to maximize the use of resources within the system among the active jobs. The proposed scheduler has been evaluated against Fair Scheduler on virtual cluster built on a physical cluster of 20 machines. The results demonstrate a gain of about 12% increase in throughput of Jobs.*

*Keywords: Cloud Computing, Data Locality, MapReduce, Virtualization, Hadoop*

## 1. INTRODUCTION

Nowadays cloud computing is expanding its services to data-intensive computing on distributed platforms like MapReduce [13], Dryad [14], and Hadoop [1]. In such distributed platforms on clouds, physical machines are virtualized, and a large variety of virtual machines (VMs) form a virtual cluster. Completion time goals of MapReduce jobs play an important role in achieving higher profits or utility for the content suppliers. Depending on whether or not a task is assigned to a node with its input data (local task or non-local task), the execution time of the task might differ considerably. Previous studies [10] [16] [17] showed that Data locality affects the throughput of Hadoop jobs considerably, even if alternative parameters have an effect on the performance of MapReduce jobs. However, to boost locality, when a task is scheduled, the computing node with the corresponding data should have free computing slots to run the task. If not the task should be assigned to a remote node which requires remote data transfer from another node for the task to run. The challenge that needs to be addressed is how efficiently we can schedule the jobs such that both data locality and deadline requirements are satisfied.

In this paper we propose and evaluate a novel scheduler that dynamically estimate the resources required by a job to meet the deadline and also schedule the jobs to maximize data locality. The contribution of this paper is twofold. Initially, a mechanism for determining the minimum number of map and Reduce slots to every job for supporting the job completion goals is presented. Job characteristics are estimated dynamically by running an initial set of tasks of the jobs. Based mostly on the task characteristics, for every job with a specified deadline, we are able to estimate and allocate the suitable range of map and Reduce slots needed for finishing the task within the deadline. The important feature of this mechanism is that as the time progresses and also the job deadline gets nearer, the introduced mechanism will re-computes the number of resources required by every job to fulfill its deadline. Another contribution is to present a way of reconfiguring the resources dynamically to realize maximum data locality through virtualized resources. For example, the number of virtual CPUs for a virtual machine may be increased or decreased dynamically while the virtual machine is running. To the best of our knowledge, the proposed scheduler is the first effort that attempt to schedule jobs to meet deadlines with increased data locality through Virtualization. The remainder of this paper is organized as follows: we give a brief overview of MapReduce and the resource estimation model for MR Jobs in section 2. Next, we present our algorithms for the proposed scheduler based on resource estimation model and dynamic resource reconfiguration in section 3. Section 4 comprises the evaluation of our proposed scheduler and summary of results. In Section 5 Related work is presented. Contributions and future work are stated in section 6.

## 2. PRELIMINARIES

### 2.1) MapReduce

The MapReduce programming model consists of 2 data processing functions: Map and Reduce. Input data is partitioned into fixed sized blocks and fed into parallel Map tasks that process the data chunks and turn out intermediate output as a set of key-value tuples. The map task applies the user-defined map performs on every record and buffers the ensuing output. This intermediate data is hash-partitioned based on key for the various Reduce tasks and written to the local disk of the worker executing the map task. Every Reduce task performs 3 steps: copy - the map output is copied to reducer nodes, sort: - the collected map output is sorted based on key and Reduce - reduce perform e.g. Aggregation is applied to the data. Jobs are composed of a collection of tasks. Job scheduling in Hadoop is performed by the Job master (Job tracker), that manages a variety of worker nodes (Task trackers) in the cluster. Each worker has a fixed number of map and reduce slots, which can run tasks. Different schedulers [12] follow different approaches to improve the performance of MapReduce Jobs. The number of map and reduce slots is statically configured (typically with one or two per core). The workers periodically send heartbeats to the master to report the number of free slots and the progress of the tasks that they are currently running. Based on the free slot availability and scheduling policy, the master assigns map and reduce tasks to slots in the cluster. The Hadoop Distributed File System (HDFS) [2] supports reliability and fault tolerance of MapReduce computation by storing and replicating the inputs and outputs of a Hadoop job.

### 2.1) Resource Estimation Model:

For users who require service guarantees, a performance question to be answered is the following: *Given a MapReduce job J to process data of size Φ and be completed within the Completion time D, how many map/reduce slots need to be allocated to this job over time so that it finishes within deadline D when run in a MapReduce cluster having N nodes with Nm map task slots, Nr reduce task slots and possibly k jobs executing at the time.* In this section, we lay the foundation for dealing with the deadline requirements in Hadoop-based data processing by proposing a resource estimation model of MapReduce jobs that dynamically estimates the completion time of a job during its execution. While doing so, we take advantage of the fact that MapReduce jobs are a collection of a large number of smaller tasks. Specifically, we hypothesize from a subset of tasks that have completed so far, we can predict the properties of remaining tasks. We acknowledge the fact that MapReduce tasks vary widely in their execution characteristics depending on the data set they process. Hence, we do not expect this estimation technique to provide accurate predictions all the time, but we do expect that when combined with dynamic scheduling it will permit fair management of completion times of multiple jobs. The technique targets a highly dynamic environment in which new jobs can be submitted at any time. Thus, the actual amount of resources available for MapReduce applications can vary over time. The proposed scheduler introduced in this paper uses the completion time estimate for each job to determine the resource demand of all jobs. The minimum unit of resource allocation is the slot, which corresponds to a worker process created by a TaskTracker.

**Table 1**: Description of symbols used in modeling MapReduce Jobs

| Symbol | Description |
| --- | --- |
| $D$ | Deadline of Job |
| $J$ | Set of Jobs |
| $n_m^j$ | Map slots allocated to job $j$ to finish |
| $n_r^j$ | Reduce slots allocated to job $j$ to finish |
| $u_m^j$ | Number of map tasks of job $j$ |
| $v_r^j$ | Number of Reduce tasks of job $j$ |
| $t_m^j$ | Time taken by a map task of job $j$ to finish |
| $d_m^j$ | Remaining time of a map task to finish. |
| $e_m^j$ | Elapsed time of a map task |
| $t_r^j$ | Time taken by a Reduce task of job $j$ to complete |
| $t_s^j$ | Time taken by a copy operation during Shuffle and sort phase (dependent on network bandwidth) |
| $C^j$ | Set of completed map tasks of Job $j$ |
| $U^j$ | Set of not yet started tasks of Job $j$ |
| $R^j$ | Set of running tasks of Job $j$ |

Each TaskTracker has a number of slots for Map tasks and another number of slots for Reduce tasks. In general in a MapReduce application the execution time is dominated by the time required to complete the Map tasks, but there are cases where the Reduce tasks also dominate. In both cases, jobs start with a Map phase, in which performance data is collected, and is followed by the Reduce phase. The scheduler cannot make assumptions about the Reduce phase before seeing some Reduce tasks completing. Therefore, in the absence of information from previous runs, the scheduler needs to estimate the effort of the Reduce phase compared to the Map phase, such that overall job execution time can be predicted before the Reduce phase starts. Table 1: describes the symbols used in modeling the MapReduce jobs. When a job is submitted for execution with a completion time goal there is no data available and it is not possible to estimate the required slots. The individual task completion time of this job is calculated based on completing tasks so far as proposed below.

Mean completed map task length of Job $j$ can be estimated as below:

$$\mu_m^j = \frac{1}{|C^j|} \sum_{m \in C^j} t_m^j \quad (1)$$

On the fly, the completion time of a map task $t_m^j$ is same as mean completion time. Therefore

$$\mu_m^j = t_m^j \quad (2)$$

We have assumed all the nodes in the cluster are homogeneous in nature, the assumption made is both map and reduce tasks will take the same amount of time to finish. Therefore

$$t_m^j = t_r^j \quad (3)$$

As each MapReduce job $j$ consists of $u_m^j$ Map tasks and $v_r^j$ Reduce tasks,
Total Map phase duration can be estimated as:

$$\frac{u_m^j \cdot t_m^j}{n_m^j} \quad (4)$$

Similarly, the Total Reduce phase duration can be estimated as:

$$\frac{v_r^j \cdot t_r^j}{n_r^j} \quad (5)$$

A MapReduce job $j$ with $u_m^j$ Map tasks and $v_r^j$ Reduce tasks require $(u_m^j \cdot v_r^j)$ distinct copy operations as part of shuffle and sort phase since each mapper may produce intermediate output going to every reducer. Therefore, total Shuffle and Sort phase duration can be estimated as:

$$(u_m^j \cdot v_r^j) t_s^j \quad (6)$$

Hence, total Completion time of the job is:

$$\frac{u_m^j \cdot t_m^j}{n_m^j} + \frac{v_r^j \cdot t_r^j}{n_r^j} + (u_m^j \cdot v_r^j) t_s^j \leq D \quad (7)$$

The above equation is rewritten to find the minimum no. of Map/Reduce slots to be allocated to a job $j$ to meet its deadline.

$$\frac{u_m^j \cdot t_m^j}{n_m^j} + \frac{v_r^j \cdot t_r^j}{n_r^j} \leq D - (u_m^j \cdot v_r^j) t_s^j \quad (8)$$

To make the calculations simple, the Equation (8) can be written as:

$$\frac{A}{n_m^j} + \frac{B}{n_r^j} = C \quad (9)$$

Where
$A = u_m^j \cdot t_m^j \; ; \; B = v_r^j \cdot t_r^j \; ; \; C = D - (u_m^j \cdot v_r^j) t_s^j$
The above equation is of the form
$f(n_m^j, n_r^j) = n_m^j + n_r^j$ over $\frac{A}{n_m^j} + \frac{B}{n_r^j} = C$

As we need to find the minimum no. of map and reduce slots to meet the deadline, now the objective is to minimize $(n_m^j, n_r^j)$. Lagrange's Multiplier method can be used to find the minima of a function $f(x, y)$ Subject to $g(x, y) = c$.
Solving the Equation (9) using Lagrange's Multiplier Method gives:

$$n_m^j = \frac{\sqrt{A}(\sqrt{A}+\sqrt{B})}{C} \quad \& \quad n_r^j = \frac{\sqrt{B}(\sqrt{A}+\sqrt{B})}{C} \quad (10)$$

Where $n_m^j$ and $n_r^j$ are the minimum numbers of Map and Reduce slots required to complete the job within the deadline.

## 3. RELATED WORK

The default scheduling policy of Hadoop is First in First out (FIFO). Under this scheme, the job that was submitted earlier gets preference over jobs submitted later. Recent efforts such as Delay Scheduler [15], Dynamic Proportional Scheduler [13] offer differentiated service for Hadoop jobs allowing users to adjust the priority levels assigned to their jobs. However, this does not guarantee that the job will be completed by a specific deadline. [11] Considers deadline constraints in the context of real time transactions in single processor environments. FLEX [6] extends HFS by proposing a special slot allocation scheme that aims to optimize explicitly some given scheduling metric. Polo et al. [8] introduced an online job completion time estimator which can be used for adjusting the resource allocations of different jobs. However, their estimator tracks the progress of the map stage alone and has no information or control over the reduce stage. Phan et al. [7] aim to build an off-line optimal schedule for a set of MapReduce jobs with given deadlines by detailed task ordering of these jobs by formulating the scheduling problem as a constraint satisfaction problem (CSP). So far significant work has been done to improve MapReduce platforms on virtualized cluster environments. Purlieus [18] improved the locality of map and reduce tasks in MapReduce platforms on the cloud by locality-aware VM placement by exploiting prior knowledge about the characteristics of MapReduce workloads Using the data layout, the VM scheduler places VMs to the physical systems with the corresponding input data. Sandholm et al [9] proposed a dynamic VM reconfiguration mechanism with resource hot-plugging, to address skewed resource usages in MapReduce task executions. They assumed that multiple MapReduce clusters share physical resources, and a cluster can use more resources than another cluster, which may violate SLAs. In addition, resource sharing between virtual clusters should occur only within a single physical machine boundary. Kang et al.[20] improved the performance of virtual MapReduce cluster by modifying the context-switching mechanism of the Xen credit scheduler for MapReduce platforms. Zaharia et al. [17] addressed the performance heterogeneity problem of MapReduce platforms on virtual clusters, where sharing I/O and network resources among customers cause performance interferences. Distributed Resource Scheduler (DRS) [4] places VMs to maximize the utilization, and live-migration of VMs across physical systems to avoid resource conflicts in a system. Our work differs from all of those as we achieved the completion goals of the jobs through dynamic adjustment of resources.

## 4. JOB SCHEDULING THROUGH VIRTUALIZATION

### 4.1) Resource Reconfiguration:

In traditional data-intensive computing on physical clusters, it is not possible to change physical resources dynamically, as each physical system has a fixed amount of physical resources. However, through Virtualization it is possible to reconfigure virtual machines constituting a virtual cluster built on a physical cluster. A data-intensive platform such as MapReduce that runs on the virtual cluster can provide a highly flexible environment, which can scale up and down accommodating changing computation demands of various users. Cloud providers consolidate virtual clusters from different users into a physical data center, to maximize the utilization of resources. Current virtual clusters for data-intensive computing can support the flexibility of selecting the type of computing nodes and the number of nodes in a cluster, when the cluster is configured. Users can choose the most appropriate virtual cluster configuration to meet their computational requirements. Although such a static configuration of each virtual machine in the cluster can still provide better flexibility than clusters with physical machines, the static configuration cannot satisfy dynamically changing computing demands during the life time of a virtual cluster. To adapt to changing demands on each VM in a virtual cluster, each VM may be reconfigured dynamically. Current Virtualization techniques can support such a dynamic reconfiguration of each virtual machine. As long as physical resources are available, each virtual machine can be assigned with more virtual CPUs and memory while the virtual machine is running. However, the currently available cloud services, such a dynamic reconfiguration of the VM is not available. The Configuration Manager (CM) component sends Assign and Release requests to the hypervisor running on each physical machine. Master node of each virtual cluster runs the CM and one CM is created for each virtual cluster. If resource reconfiguration is needed for a Virtual Machine, CM sends an Assign or Release request to Machine Manager (MM) running in the physical system, where the VM to be reconfigured is running on. After receiving requests from CM, MM re-assigns the requested virtual CPUs to the VM. This is done by the Machine Manager through the requests made to the hypervisor of the physical machine, to hot-plug or to un-plug virtual CPUs for VMs running on the system. As outlined in Algorithm 1, to maximize data locality, a task is delayed until a core becomes available in the target node. Releasing and assigning cores in the source and target VMs are done in decoupled manner. If a VM has a free slot, it registers the free core to the Release queue (RQ) of the system. If a

VM has a pending local task assigned by RC, the task is appended to the Assign queue (AQ) of the system. As soon as both the AQ and RQ of the same system (MM) has at least an entry, VM reconfigurations occur in the system, Releasing a core from a VM, and assigning a core to another VM in the same system.

---

**Algorithm 1. Map Task Assignment through Resource reconfiguration Algorithm**

**Input:** job *j* and node *n*

1: **if** job j has unassigned local map task $t_i^j$ **then**
2:     Launch local task $t_i^j$ on node *n*
3: **else if** j has unassigned non local map task $t_i^j$ **then**
4:     $S_{rq}$ = {set of nodes storing data of $t_i^j$ in the descending order of number of entries in Release Queue}
5:     **if** $S_{rq}$ is not empty **then**
6:         $p := S_{rq}[top]$
7:     **else**
8:         $S_{aq}$ = {set of nodes storing data of $t_i^j$ in the ascending order of number of entries in Assign Queue.}
9:         $p := S_{aq}[top]$
10:    **end if**
11:    add an entry to Assign Queue of P*p*.
12:    add an entry to Release Queue of P*n*.
13:    Launch map task $t_i^j$ on node *p*.
14: **end if**

---

This delayed task scheduling occurs since the CPU resource cannot be transferred beyond the physical system boundary directly. Even if the system running the source VM has a free core, the computing resource cannot be directly available to the target VM. However, with multiple VMs sharing a physical system, the target system will soon have a free core, as a task finishes in one of the VMs, and a local task is not found for the VM. In this approach queuing delay can be an important factor for cluster performance, as resource utilization might be degraded if handling request of the assign and release queue is postponed due to a large queuing delay.

**4.2) Completion time based Scheduler**

Based on the Resource modeling introduced in section 2 and dynamic map task assignment through Resource reconfiguration, a novel scheduler is presented in this section. Algorithm 2 outlines the scheduling policy. After a job is submitted, there is no data available and it is not possible to estimate the required slots or the completion time. Therefore, jobs with no completed or running tasks always take precedence over other jobs. If there is more than one such job, the oldest one comes first. The first component of the scheduler is to estimate the minimum number of Map/Reduce slots required to complete a job within the deadline. Initially as no data is available for a job, individual jobs are executed alone to obtain the estimate of task completion time. Then the minimum number of Map/Reduce slots required by a job to complete within the deadline is estimated. TaskTracker nodes will update their status to the JobTracker periodically or when it has finished its task using heartbeat mechanism. Usually the heartbeat interval is 3s. Data locality is more crucial in the map phase as more than one map tasks of a job can execute in parallel with different input splits. Where as same is not the case with reduce phase because each reducer gets the input almost from all the mappers. Due to this, in the proposed scheduler we have considered only the map phase to maximize data locality. The scheduler also decides whether to consider the job for execution based on the running tasks of the jobs (line 7 & 10). If the running tasks are more than the minimum slot requirement, next job is considered for scheduling. To support dynamic VM reconfiguration for Hadoop platforms, the cloud provider must offer cluster-level pricing options to allow users to choose the base VM configuration and the number of VMs. However, the configuration of a VM can be dynamically adjusted with more or less virtual CPUs than the base VM configuration, while the total cores assigned to the cluster does not change. Virtual core to VMs may be added or removed to maximize data locality.

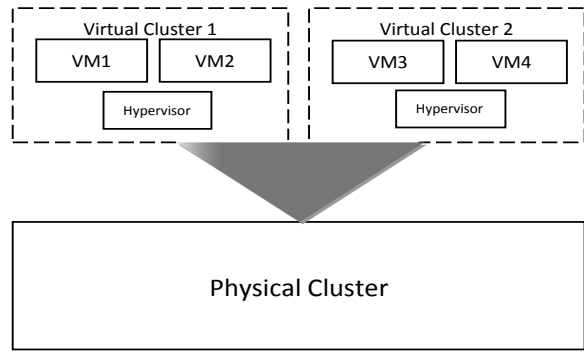

Figure 1: Virtual Cluster on a Physical Cluster

**Algorithm 2: Completion time based Scheduling Algorithm**

1: When job j is added:
2: Calculate the minimum number of Map and Reduce slots ($n_m^j, n_r^j$) required by Job j using Equation (10)
3: When a heartbeat is received from node *n*:
4: **if** node *n* has a free slot **then**
5:    Sort jobs in the ascending order of their deadlines
6:       **for** each job j in jobs **do**
7:    **if** !j.mapfinished and j.ScheduledMaptasks <$n_m^j$ **then**
8:          *taskassignment (j,n);*
9:    **end if**
10: **if** j.mapfinished and j.ScheduledReducetasks < $n_r^j$ **then**
11:       **if** job j has unassigned reduce task $t_i^j$ **then**
12:            Launch task $t_i^j$ on node *n*
13:       **end if**
14:    **end if**
15:    **end for**
16: **end if**
17: **for** each task $t_i^j$ finished **do**
18:    $C^j = \{C^j\} \cup t_i^j$
19: Re compute the number of Map and Reduce slots ($n_m^j, n_r^j$) required by Job j based on current progress using Equation (10)
20: **end for**

---------------------------------------------------------------------------------------------------------------------------------

## 5. EVALUATION

We have conducted the experiments to evaluate the proposed scheduler against Hadoop Fair scheduler [3]. We have chosen five different kinds of MapReduce jobs for experimental evaluation. We have simulated the Datacenter environment through a cluster of 20 physical machines. Each node is configured to have two map slots and two reduce slots. Xen 4.0.1 is used to virtualize physical machines. Xen credit scheduler [5] is used to reconfigure Virtual machines which supports virtual CPU hot-plugging. Hadoop 0.20.2 is used for executing MapReduce jobs. We have evaluated the proposed scheduler against Fair scheduler with five different jobs with different input data sizes. Namely:

*1) Word Count:* The 'Word Count' application is one of the sample applications contained in the Hadoop distribution. It takes a set of text files as input, and counts the number of times that each word appears.
*2) Sort:* This application sorts a set of records that is randomly generated. The application uses identity map and identity reduce functions as the MapReduce framework does the sorting.
*3) Grep: This* application generates intermediate data simply indicating if a word occurs in input data. These kinds of applications generate small intermediate data.
*4) Permutation Generator:* This application generates permutations of input strings and is reduce-input heavy workload as it generates large amount of intermediate data for the reducers.
*5) Inverted Index:* This application generates a sequence of words and the list of documents that contains the word. The set of all output pairs forms a simple inverted index. The experiments were conducted in two phases, first phase with Fair scheduler and the second phase with the proposed scheduler.

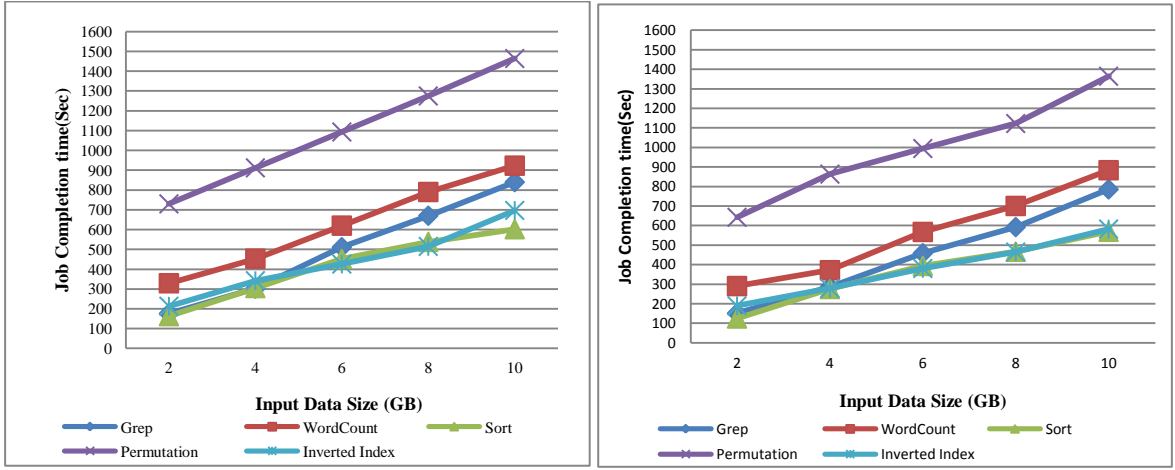

(a) With Fair Scheduler            (b) With Proposed Scheduler

Figure 2: Job Completion times of various with different input data

All the above jobs are executed withthe same input dataa sizes of 2GB, 4GB, 6GB, 8GB and 10GB. We have conducted the same set of experiments with the same input data using Fair scheduler and the proposed scheduler. The results are plotted in figure 2 (a) and figure 2 (b). It is observed from the results that as the input data size increase, the completion time of jobs will also increase. The completion times will change based on the type of workload as different workloads have different resource demands. For example, the Permutation generator job results in more intermediate output data, which require huge number of copy operations in shuffle phase leading to increase in completion time relatively more than the other workloads. We can observe from the results that the reduction of completion times of jobs with the proposed scheduler over fair scheduler. In another experiment, random input sizes were considered for all the jobs and Algorithm 2 outlines the scheduling policy. After a job is submitted, there is no data available and it is not possible minimum slots required were calculated based on the equation (10). Table 2: Shows the deadlines of jobs along with a minimum number of slots required to complete the job within the deadline. This set of jobs is executed with both the Fair scheduler and the proposed scheduler. The results show an improvement of throughput of jobs over fair scheduler. Through the results we can conclude that the proposed scheduler minimizes the completion times by executing more data local tasks. From figure 3 it is evident that the completion times of permutation generator job both with the fair and proposed scheduler is almost same as this job is a reduce input heavy job which require a large number of copy operations during the shuffle phase. Data locality is less significant in reduce phase as the intermediate data need to be copied from all the mappers to all reducers. It does not make sense to launch a data local task. For other jobs the proposed scheduler significantly reduces the completion time against fair scheduler. As the jobs are concurrently running it is obvious that a VM may not have free slots to launch a data local task. Hence tasks have to wait for the data local VMs to become free. This wait time also needs to be considered in estimating the throughput of jobs. We observed from the experiments that this wait time is negligible because the tasks of MapReduce jobs will be finished in less than a minute to release slots of a VM for other jobs. Through the evaluation of proposed scheduler against the fair scheduler we can conclude that proposed scheduler result in the gain of about 12% of throughput of MapReduce jobs on a virtual cluster.

**Table 2:** Slot Allocation to different types of jobs with Completion time goals

| Job Type | Job Completion time (Sec) | Input Size (GB) | Slots Required to meet Completion time | |
|---|---|---|---|---|
| | | | Map Slots | Reduce Slots |
| Grep (Word Search) | 650 | 10 | 24 | 8 |
| Word Count | 520 | 5 | 14 | 7 |
| Sort | 500 | 10 | 20 | 11 |
| Permutation Generator | 850 | 4 | 15 | 16 |
| Inverted Index | 720 | 8 | 12 | 9 |

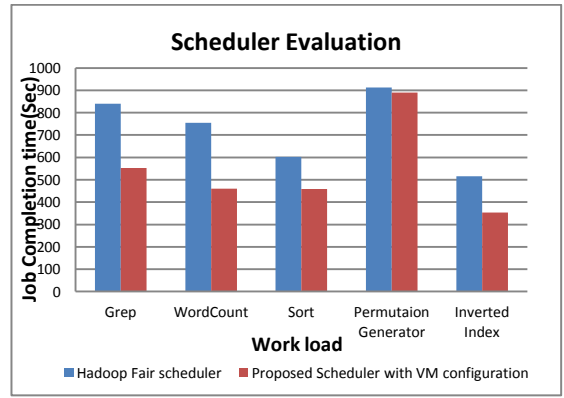

Figure 3: Comparison of Job completion times

## 6. CONCLUSION AND FUTURE WORK

Many of the applications on the cloud need to achieve the performance goals such as deadlines. In this work, we proposed and evaluated a novel scheduler that extended the real time cluster scheduling approach to derive minimum map and reduce task slot count for performing task scheduling with deadline constraints in Hadoop. Our current job ordering is inspired by the EDF scheduling. The proposed scheduler improves the input data locality of a virtual MapReduce cluster, by temporarily increasing cores to VMs to run local tasks. With VM reconfiguration, each node can be adjusted to provide only the necessary amount of resource demanded for that node. Addressing resource requirements of different types of workloads (CPU/IO) which may cause load imbalance and under utilization of the cluster will be our future work